\documentclass[aps,prx,twocolumn,superscriptaddress,showpacs]{revtex4-1}
\usepackage{amsmath,amssymb,graphics,epsfig,epstopdf,color,verbatim,ulem,braket,tabularx}
\usepackage{multirow}
\usepackage[colorlinks,linkcolor=blue,citecolor=blue,urlcolor=blue,bookmarks=false]{hyperref}

\usepackage{listings}
\usepackage{cancel}
\usepackage{mathrsfs}
\usepackage{soul}
\usepackage{color}
\usepackage{url}

\begin{document}

\title{Possible Enhancement of Superconductivity in Ambient-Pressure La$_3$Ni$_2$O$_7$ Thin Film}

\author{Yichen Hua}
\affiliation{Department of Physics, Southern University of Science and Technology, Shenzhen 518055, China}
\affiliation{Quantum Science Center of Guangdong-Hong Kong-Macao Greater Bay Area, Shenzhen 518045, China}

\author{Wenxin He}
\affiliation{Department of Physics, Southern University of Science and Technology, Shenzhen 518055, China}

\author{Wei-Qiang Chen}
\affiliation{State Key Laboratory of Quantum Functional Materials, Department of Physics, and Guangdong Basic Research Center of Excellence for Quantum Science, Southern University of Science and Technology (SUSTech), Shenzhen 518055, China}

\author{Jian-Jian Miao}
\email{miaojianjian@quantumsc.cn}
\affiliation{Quantum Science Center of Guangdong-Hong Kong-Macao Greater Bay Area, Shenzhen 518045, China}
 
\author{Changming Yue}
\email{yuecm@sustech.edu.cn}
\affiliation{State Key Laboratory of Quantum Functional Materials, Department of Physics, and Guangdong Basic Research Center of Excellence for Quantum Science, Southern University of Science and Technology (SUSTech), Shenzhen 518055, China}

\begin{abstract}
As an unconventional superconducting system capable of reaching 60 K under ambient pressure, the La$_3$Ni$_2$O$_7$ thin film superconductor has recently become a focal point in the field of superconductivity, calling for further theoretical exploration of its possible pairing mechanisms. In this work, we employ the fluctuation exchange (FLEX) approximation to systematically analyze the superconducting properties of a previously proposed two-site, two-orbital model for the La$_3$Ni$_2$O$_7$ thin film in the weakly correlated regime, with particular emphasis on its dependence on hole doping. Through a more detailed examination of the Fermi-surface topology, we find that when a $\delta$ pocket composed of the $d_{z^{2}}$ antibonding orbital emerges near the $\Gamma$ point, its nesting with the $\gamma$ pocket, together with the nesting between the $\alpha$ and $\beta$ pockets, leads to a mutual enhancement of $s{\pm}$-wave pairing at the corresponding wave vector. Furthermore, we propose that this nesting-driven enhancement of spin-fluctuation-induced pairing may provide a viable mechanism for enhancing superconductivity.
\end{abstract}
\maketitle


\newpage

\section{Introduction.} \ 

The mechanism of high-temperature superconductivity has long been a central research topic in condensed matter physics. In recent years, following the discovery of cuprate and iron-based superconductors, the emergence of the nickel-based superconductor family \cite{LDF2019,Wang327_Nature2023} has attracted wide attention. Among them, Ruddlesden-Popper (RP) phase nickel-based superconductors, exemplified by the La$_{3}$Ni$_{2}$O$_{7}$ compound, can achieve a superconducting transition temperature of up to 80K under high pressure conditions \cite{Wang327_Nature2023}. During the past two years, significant progress has been made in both experimental characterization \cite{hou2023emergence,Zhang_2024_HighTc,Xie_2024_MagneticExchange,Li_2024_ElectronicCorrelation,Wang_2024_La2PrNi2O7_Nature,dan_2024,wang2024pressure,cai_2024,liu_2024,ren2025resolving} and theoretical studies \cite{YaoPRL2023,Yang_2023_sPlusMinus,Liu_2023_sPlusMinus,Yang_2023_ValenceBonds,Qin_2023_HighTc,zhang_2023,pupha_l2023,Christiansson_2023,pupha_l2023,liao2023electron,liao_2024,zhang2024electronic,zhang2023trends,Tian_2024_Correlation,Oh_2024_StrongPairing,Fan_2024_BilayerCoupling,RyeePRL2024CMDFT,wang2024electronic,chen2024electronic,lu2024interlayer,sakakibara2024,jiang2024high,zhang2024strong,Zheng_2025_TwoOrbital,jiang2025theory,duan2025orbital,xi2025transition,shi2025theoretical,ouyang2025two,maier2025interlayer,zhan2025cooperation,le2025opposite,wang2025electronic,gu2025effective,bejas2025out,puphal2025superconductivity,xia2025sensitive,zhang2025mirror,jiang2025dual}
of high-pressure nickel-based superconductors. One of the central questions in the theoretical study of nickel-based superconductors is how to further enhance their superconductivity, enabling them to operate at lower pressures and higher temperatures. The shape of the Fermi surface plays a critical role in determining superconducting behavior \cite{Wang327_Nature2023,Yang_2023_sPlusMinus,Liu_2023_sPlusMinus,Yang_2023_ValenceBonds}. Importantly, the Fermi surface is sensitive to variations in lattice constants, electron filling, and interaction strength \cite{wu2024ac,jiang2024high,huo2025modulation,wang2025electronic}. Theoretically, many research groups have explored various strategies such as chemical doping, applying pressures to tune the Fermi surface geometry and the density of states at the Fermi level, trying to  enhance superconductivity \cite{wang2023observation,wu2024ac,rhodes2024structural,geisler2024structural,chen2025charge,huo2025modulation}. 

Recently, the research team led by Z.Y.Chen achieved a breakthrough by successfully growing a La$_{2.85}$Pr$_{0.15}$Ni$_2$O$_7$ thin-film superconductor on an SrLaAlO$_{4}$ substrate under ambient pressure, with a superconducting transition temperature reaching about 40K \cite{Ko_LNO327AmbientSC,Zhou_2024_AmbientSC}. Subsequently, by refining the growth techniques, they further increased the superconducting transition temperature to 60 K \cite{zhou2025superconductivity} stimulating extensive experimental and theoretical discussions regarding their superconducting mechanism \cite{liu2025superconductivity,XJZhou_ARPES_327,li2025photoemission,ryee2025superconductivity,ushio2025theoretical,shi2025effect,yue2025correlated,gao2025theoretical,wang2026superconducting}. In particular, the role of the Fermi-surface structure in superconductivity seems to have emerged as a key topic in recent discussions \cite{ryee2025superconductivity,ushio2025theoretical,shi2025effect,gao2025theoretical}.
The theoretical study by the authors  \cite{yue2025correlated} reveal this complex heterostructure is a moderately correlated system far from doped Mott insulator regime. Our results suggested that the superconductivity in ambient pressure  La$_3$Ni$_2$O$_7$ films may originate from spin-fluctuation-induced $s_{\pm}$-wave pairing.

However, the significance of the ambient-pressure La$_3$Ni$_2$O$_7$ thin film superconductor extends far beyond this single material. Its study may not only guide the discovery of nickel-based superconductors with higher superconducting transition temperatures under ambient conditions, but also suggests that the structure of La$_3$Ni$_2$O$_7$ thin films can be tuned by selecting different substrates \cite{osada2025strain}. This, to some extent, alleviates the experimental difficulty of doping in nickel-based systems and provides a new platform for gaining deeper insights into the superconducting pairing mechanism \cite{hao2025superconductivity,wang2025recent}.

We find that, compared with the bulk La$_3$Ni$_2$O$_7$ compound, the La$_{2.85}$Pr$_{0.15}$Ni$_2$O$_7$ thin-film structure exhibits a larger inter-layer Ni–Ni distance within each bilayer, which directly reduces the inter-layer hopping $t_\perp^z$ between the $d_{z^2}$ orbitals  \cite{yue2025correlated}. A smaller $t_\perp^z$ brings the bonding and anti-bonding states of the $d_{z^{2}}$ orbital closer in energy and induces a tendency toward the emergence of a new Fermi surface. If the $c$-axis lattice constant can be further expanded experimentally, it would likely be favorable for enhancing the superconductivity of nickel-based systems.

In this work, we employ the fluctuation exchange (FLEX) approximation to investigate the La$_{3}$Ni$_{2}$O$_{7}$ thin film system in the weak coupling regime. By analyzing the Fermi surface (FS) nesting mechanism, we predicted a possible scenario that could enhance superconductivity in this system. Specifically, we first discuss the key features of the FLEX-corrected Fermi surface and spin correlation functions. We then compute and analyze the leading eigenvalue $\lambda$ of the linearized Eliashberg equation and its corresponding gap function on the Fermi surface, exploring possible enhance mechanisms of pairing based on the numerical behavior of $\lambda$. Finally, we summarize and discuss our findings.

\section{Method} \ 

\begin{figure}
\includegraphics[clip,width=3.4in,angle=0]{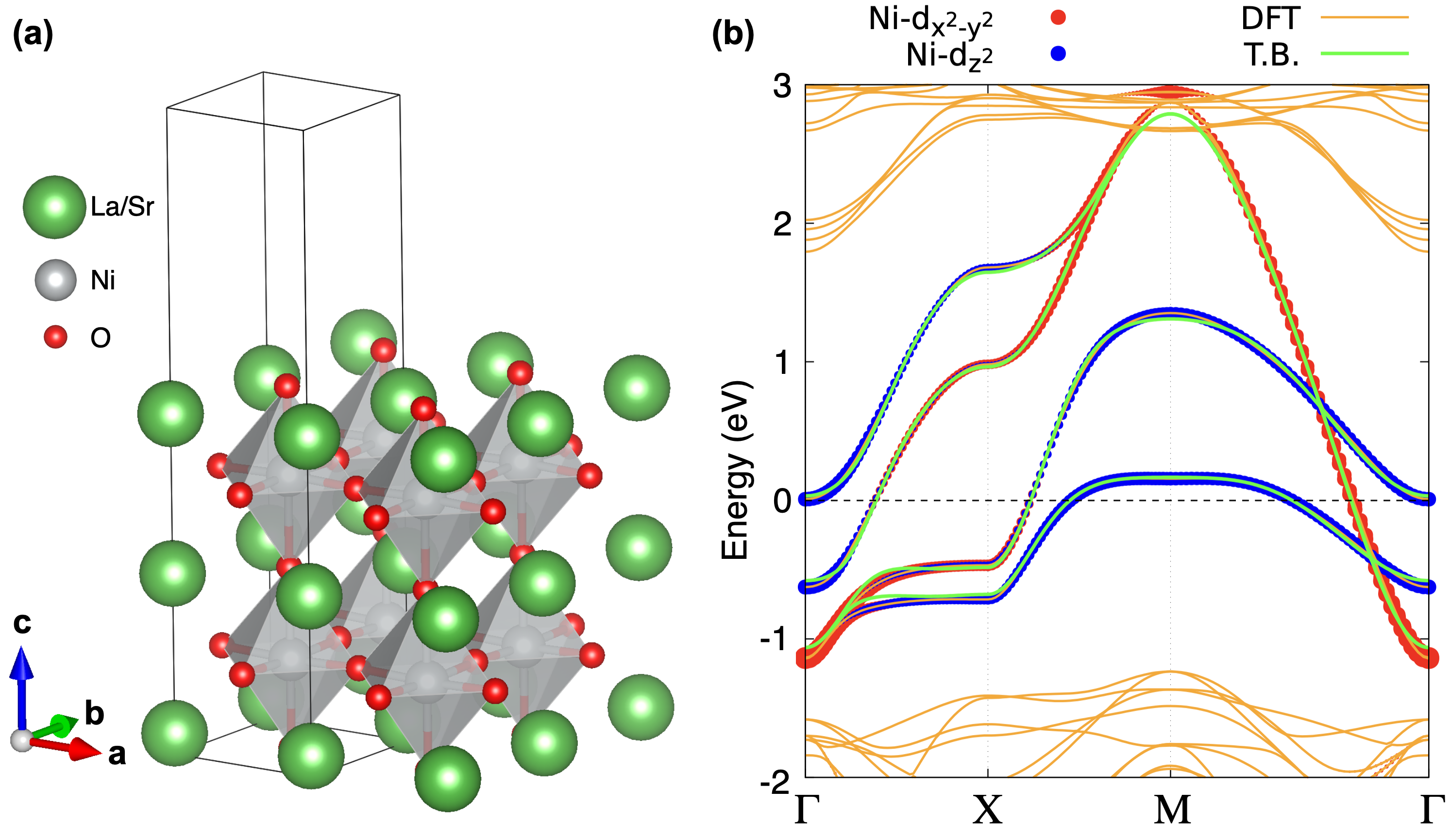}
\caption{(a), The schematic crystal structure of half-UC thin film of La$_3$Ni$_2$O$_7$ using the lattice parameters of La$_{2.85}$Pr$_{0.15}$Ni$_2$O$_7$ grown on SrLaALO$_4$ substrate \cite{Zhou_2024_AmbientSC}, where nickel atoms form octahedral coordination with surrounding oxygen atoms. For a better illustration, we put a exaggeratedly small vacuum than the one used in calculation above the thin film. (b), The band structure of the half-UC system. The orange lines show the DFT bands, green lines the TB bands. We also show the projected bands of the $e_g$ orbitals, with the red (blue) dots the $d_{x^2-y^2}$ ($d_{z^2}$) orbital.}
\label{fig1}
\end{figure}

In this work, we simplify the complicated La$_{2.85}$Pr$_{0.15}$Ni$_2$O$_7$/SrLaAlO$_4$ heterostructure to an half-unit-cell (half-UC)  La$_3$Ni$_2$O$_7$ with the experimental parameters, including the lattice constants, atomic spacing (see Ref.~\cite{yue2025correlated}). The minimal two-site two orbital ($d_{x^{2}-y^{2}}$ and $d_{z^{2}}$ orbitals per Ni sites) model Hamiltonian is also adopted. We use $A,B$ to label the up layer and down layer and $x,z$ to label the orbital $d_{x^{2}-y^{2}}$ and $d_{z^{2}}$. So, the orbital basis in site representation can be expressed as $\Psi=[Ax,Az,Bx,Bz]^{T}$. The same Hamiltonian as in previous studies is employed \cite{yue2025correlated}, which can be written in momentum space, reads
\begin{equation}
H=H_{tb} + H_{int},
\end{equation}
where the tight-binding (TB) model part reads
\begin{equation}
H_{tb}(k)=\begin{pmatrix}
T_{k}^{x}&V_{k}&T^{'x}_{k}&V_{k}^{'} \\
V_{k}&T_{k}^{z}&V_{k}^{'}&T^{'z}_{k} \\
T^{'x}_{k}&V_{k}^{'}&T_{k}^{x}&V_{k} \\
V_{k}^{'}&T^{'z}_{k}&V_{k}&T_{k}^{z}
\end{pmatrix}.
\end{equation}
Here, 
\begin{align}
T_{k}^{x/z}&=2t_{1}^{x/z}(\cos k_{x}+\cos k_{y}) + 2t_{4}^{x/z}(\cos2k_{x}+\cos2k_{y})\nonumber\\
&+2t_{5}^{x/z}(\cos3k_{x}+\cos3k_{y})+4t_{2}^{x/z}\cos k_{x}\cos k_{y}\nonumber\\
&+\epsilon_{x/z},
\end{align}
\begin{equation}
T^{'x/z}_{k}=t_{\perp}^{x/z}+2t_{3}^{x/z}(\cos k_{x}+\cos k_{y}),
\end{equation}
\begin{equation}
V_{k}=2t_{3}^{xz}(\cos k_{x}-\cos k_{y})+2t_{5}^{xz}(\cos2k_{x}-\cos2k_{y}),
\end{equation}
and
\begin{equation}
V_{k}^{'}=2t_{4}^{xz}(\cos k_{x}-\cos k_{y}),
\end{equation}
The hopping parameters can be found in Table 2 of our previous work \cite{yue2025correlated}. The TB band structure is shown by the green lines in Fig.~\ref{fig1}(b), which agrees well with the DFT one (orange line). The interaction Hamiltonian takes the Hubbard-Kanamori form, which  reads
\begin{equation}
\begin{aligned}
H_{\text{int}} ={}&
U\sum_{i,\mu,\ell}
n_{i\mu\ell\uparrow} n_{i\mu\ell\downarrow}
+ U' \sum_{i,\mu<\nu,\sigma,\ell}
n_{i\mu\ell\sigma} n_{i\nu\ell\bar{\sigma}}
\\
&+ (U'-J)\sum_{i,\mu<\nu,\sigma,\ell}
n_{i\mu\ell\sigma} n_{i\nu\ell\sigma}
\\
&- J\sum_{i,\mu\neq\nu,\ell}
c^{\dagger}_{i\mu\ell\uparrow}
c_{i\mu\ell\downarrow}
c^{\dagger}_{i\nu\ell\downarrow}
c_{i\nu\ell\uparrow}
\\
&+ J\sum_{i,\mu\neq\nu,\ell}
c^{\dagger}_{i\mu\ell\uparrow}
c^{\dagger}_{i\mu\ell\downarrow}
c_{i\nu\ell\downarrow}
c_{i\nu\ell\uparrow}.
\end{aligned}
\end{equation}
Here, $U$ ($U^{\prime}$) is the onsite intra-orbital (inter-orbital) Coulomb interaction, and $J$ the onsite Hund coupling and pair hopping\textcolor{black}{. $i$ labels the unit cell, $\ell=A,B$ denotes the layer index, $\mu,\nu=x,z$ are orbital indices, and $\sigma=\uparrow,\downarrow$ is the spin index}. We assume a conventional choice for  $U^{\prime}$ with $U^{\prime}=U-2J$. Furthermore, $J=U/6$ is chosen throughout this work.

\begin{figure*}[htp]
    \includegraphics[clip,width=0.6\paperwidth,angle=0]{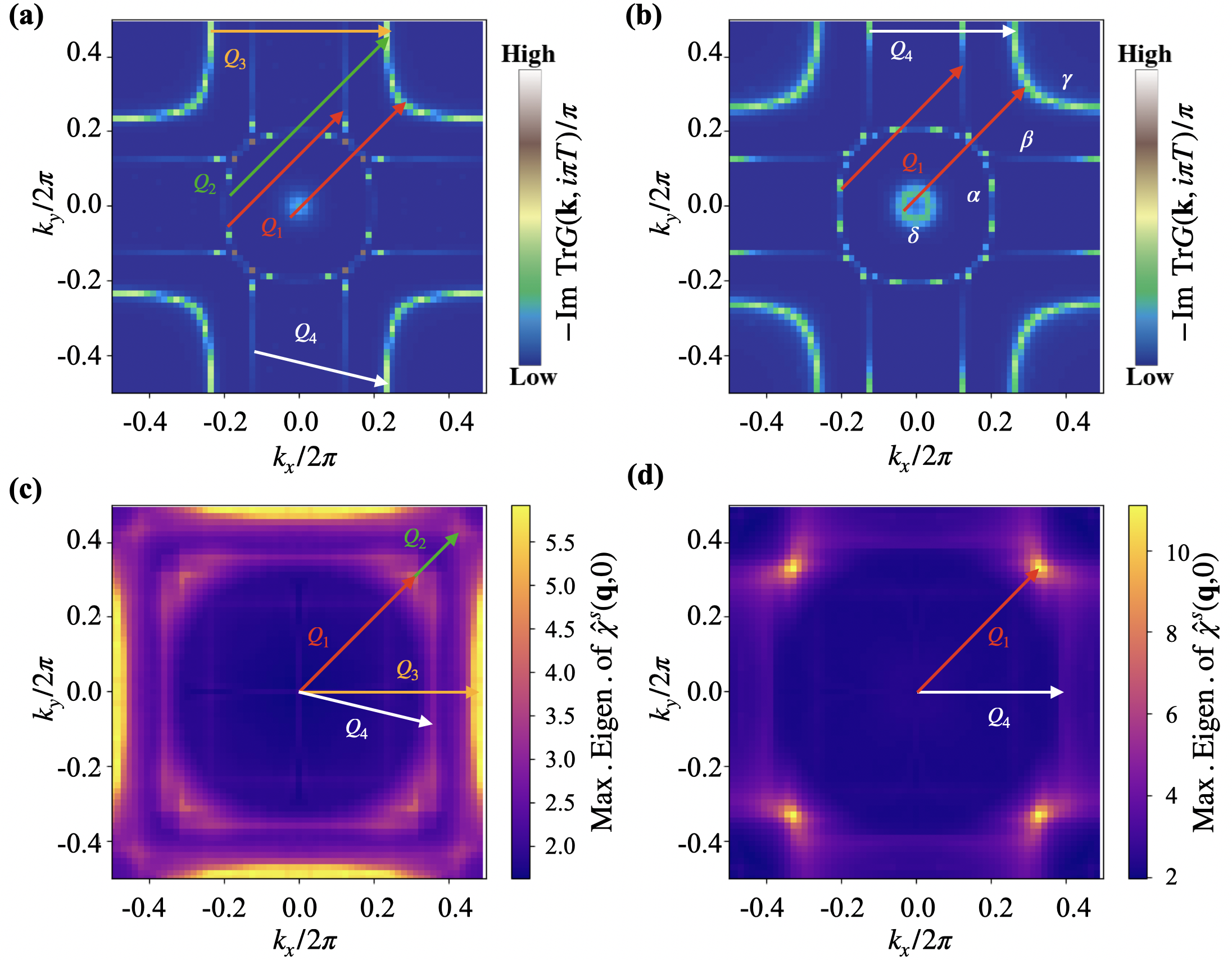}
    \caption{\textcolor{black}{(a),(b) $-\mathrm{Im}\,\mathrm{Tr}\,\hat{G}(\mathbf{k},i\omega_0)/\pi$, with $\omega_0=\pi T$, for $n=1.3$ and $n=1.42$, respectively. 
The color scale in (a),(b) is normalized by the maximum value in each panel. 
(c),(d) Largest eigenvalue of the spin-susceptibility matrix 
$\hat{\chi}^{S}(\mathbf{q})$ for $n=1.3$ and $n=1.42$, respectively. 
The four Fermi pockets, denoted by $\delta$, $\alpha$, $\beta$, and $\gamma$, are labeled in (b). 
The corresponding nesting wave vectors are marked in the relevant panels.} \textcolor{black}{The dominant nesting vectors are approximately $\mathbf{Q}_{\rm max}^{n=1.3}\simeq(0,\pi)$ and $\mathbf{Q}_{\rm max}^{n=1.3}\simeq(\pi,0)$ in (c) and $\mathbf{Q}_{\rm max}^{n=1.42}\simeq(0.65\pi,0.65\pi)$ in (d).}}
\label{fig2}
\end{figure*}

FLEX is a conserving quantum many-body approximation method that self-consistently incorporates spin and charge fluctuations by summing infinite-order bubble and ladder diagrams in the self-energy calculation \cite{bickers1989conserving,bickers1991conserving,yanase2003theory,kuroki2008unconventional,witt2021efficient,kang2011spin,bjornson2021orbital}. 
While FLEX is a perturbative approach, it has been proven highly effective in capturing the pairing symmetry driven by Fermi surface nesting.
In this work, we employ the FLEX method to calculate the band renormalization of half-UC film  La$_3$Ni$_2$O$_7$ system induced by interactions within the weak coupling regime. Here, only the particle-hole channel is considered. Its effective interaction is given by
\begin{equation}
\begin{aligned}
\hat{V}(q) ={}&
\frac{3}{2}\hat{U}^{S}\hat{\chi}^{S}(q)\hat{U}^{S}
+ \frac{1}{2}\hat{U}^{C}\hat{\chi}^{C}(q)\hat{U}^{C}
- \frac{3}{4}\hat{U}^{S}\hat{\chi}^{0}(q)\hat{U}^{S}
\\
&- \frac{1}{4}\hat{U}^{C}\hat{\chi}^{0}(q)\hat{U}^{C}
+ \frac{3}{2}\hat{U}^{S}
- \frac{1}{2}\hat{U}^{C}.
\end{aligned}
\end{equation}
\textcolor{black}{Here and hereafter, the hat denotes matrices in the composite two-particle index space $(ij)$ and $(kl)$, where each single-particle index $i,j,k,l$ runs over the basis $[Ax,Az,Bx,Bz]$. Before giving the explicit form of the interaction matrices, we write each single-particle index as a composite index. $i=(\ell_i,\mu_i)$, where $\ell_i=A,B$ denotes the layer and $\mu_i=x,z$ denotes the orbital. The interaction matrix elements are nonzero only when all four indices belong to the same layer.}  
\begin{equation}
U^{S}_{ijkl} =
\begin{cases}
U, 
& \ell_i=\ell_j=\ell_k=\ell_l,\quad
  \mu_i=\mu_j=\mu_k=\mu_l,  \\

U', 
& \ell_i=\ell_j=\ell_k=\ell_l,\quad
  \mu_i=\mu_k\neq \mu_j=\mu_l, \\

J, 
& \ell_i=\ell_j=\ell_k=\ell_l,\quad
  \mu_i=\mu_j\neq \mu_k=\mu_l,  \\

J', 
& \ell_i=\ell_j=\ell_k=\ell_l,\quad
  \mu_i=\mu_l\neq \mu_j=\mu_k, \\

0,
& \text{otherwise}.
\end{cases}
\end{equation}
\begin{equation}
U^{C}_{ijkl} =
\begin{cases}
U, 
& \ell_i=\ell_j=\ell_k=\ell_l,\quad
  \mu_i=\mu_j=\mu_k=\mu_l,  \\

-U'+J, 
& \ell_i=\ell_j=\ell_k=\ell_l,\quad
  \mu_i=\mu_k\neq \mu_j=\mu_l, \\

2U'-J, 
& \ell_i=\ell_j=\ell_k=\ell_l,\quad
  \mu_i=\mu_j\neq \mu_k=\mu_l,  \\

J', 
& \ell_i=\ell_j=\ell_k=\ell_l,\quad
  \mu_i=\mu_l\neq \mu_j=\mu_k, \\

0,
& \text{otherwise}.
\end{cases}
\end{equation}
Please notice that only intra-layer elements are non-zero. Here, $\hat{\chi}^{0}(q)$ is the bare susceptibility that reads $\chi^{0}_{ijkl}(q)=-\frac{T}{N}\sum_{k}G_{ki}(k)G_{jl}(k)$ with $G_{ij}(k)$ the Green's function calculated by Dyson's equation. $\hat{\chi}^{S}(q),\hat{\chi}^{C}(q)$ are the spin and charge susceptibilities, which can be calculated by
\begin{equation}
\hat{\chi}^{C/S}(q)
=
\left[
\hat{1}\pm \hat{U}^{C/S}\hat{\chi}^{0}(q)
\right]^{-1}
\hat{\chi}^{0}(q).
\end{equation}
In this work, we use a modified code of FLEX by Witt \cite{shinaoka2017compressing,wallerberger2023sparse,witt2021efficient}. In our calculations, we adopt a k-grid as large as $64\times64$ , and the temperature is fixed at $T = 0.001eV$. 
To ensure the validity of the FLEX approach, we restrict our discussion to cases where $U$ is relatively small. After achieving self-consistent convergence in FLEX, the obtained physical quantities are used to solve the linearized Eliashberg equation to extract information about superconducting pairing. The self-consistent gap equation to be solved is written as
\begin{equation}
\lambda\Delta_{il}(k)=\frac{T}{N}\sum_{q}\sum_{jk}V^{(\text{s}/\text{t})}_{ijkl}(q)F_{jk}(k-q)
\end{equation}
where $\hat{F}(k)=-\hat{G}(k)\hat{\Delta}(k)\hat{G}^{T}(-k)$ is the anomalous Green’s function, \textcolor{black}{and the $\text{s}(\text{t})$ means spin singlet(triplet)}. The closer the eigenvalue $\lambda$ of the self-consistent equation approaches $1$, the larger the instability towards the superconducting state with a certain pairing symmetry. In previous works, we found that pairing in this system is always predominantly spin-singlet \cite{yue2025correlated}. \textcolor{black}{Therefore, in this paper, we focus on the $s$- and $d$-wave spin singlet pairing.} 
\textcolor{black}{In the spin-singlet channel, the effective pairing interaction in the same matrix notation is given by}
\begin{equation}
\hat{V}^{\text{s}}(q)
=
\frac{3}{2}\hat{U}^{S}\hat{\chi}^{S}(q)\hat{U}^{S}
-\frac{1}{2}\hat{U}^{C}\hat{\chi}^{C}(q)\hat{U}^{C}
+\frac{1}{4}\left(3\hat{U}^{S}+\hat{U}^{C}\right).
\end{equation}

The largest eigenvalue of the linearized Eliashberg equation is efficiently computed using the power method. \textcolor{black}{To compare different pairing symmetries, we do not compute a full set of subleading eigenvalues from a single unrestricted power iteration. Instead, we solve the linearized Eliashberg equation separately in selected symmetry sectors by choosing symmetry-adapted initial gap functions. The $s$-wave calculation is initialized with a constant form factor, whereas the $d_{xy}$-wave calculation is initialized with a $d_{xy}$-symmetric form factor. The orbital/layer components are allowed to evolve freely during the iteration. Since the pairing kernel respects the lattice point-group symmetry, different irreducible representations do not mix. The reported $\lambda$ are thus the leading eigenvalues within their respective symmetry sectors. After convergence of the linearized Eliashberg equation, we project the gap function onto the bare band basis using the unitary matrix that diagonalizesthe noninteracting DFT-derived tight-binding Hamiltonian. This projection is performed only for visualization of the gap structure, the Eliashberg kernel and the Fermi-surface spectral weight are obtained from the FLEX-renormalized Green's function.} The pairing symmetry is analized from the gap function $\Delta(\mathbf{k,i\omega_{0}})$ of this system, where $\omega_{0}$=$\pi T$ is the lowest Matsubara fermi frequencies. 

\begin{figure*}[htp]
    \includegraphics[clip,width=0.8\paperwidth,angle=0]{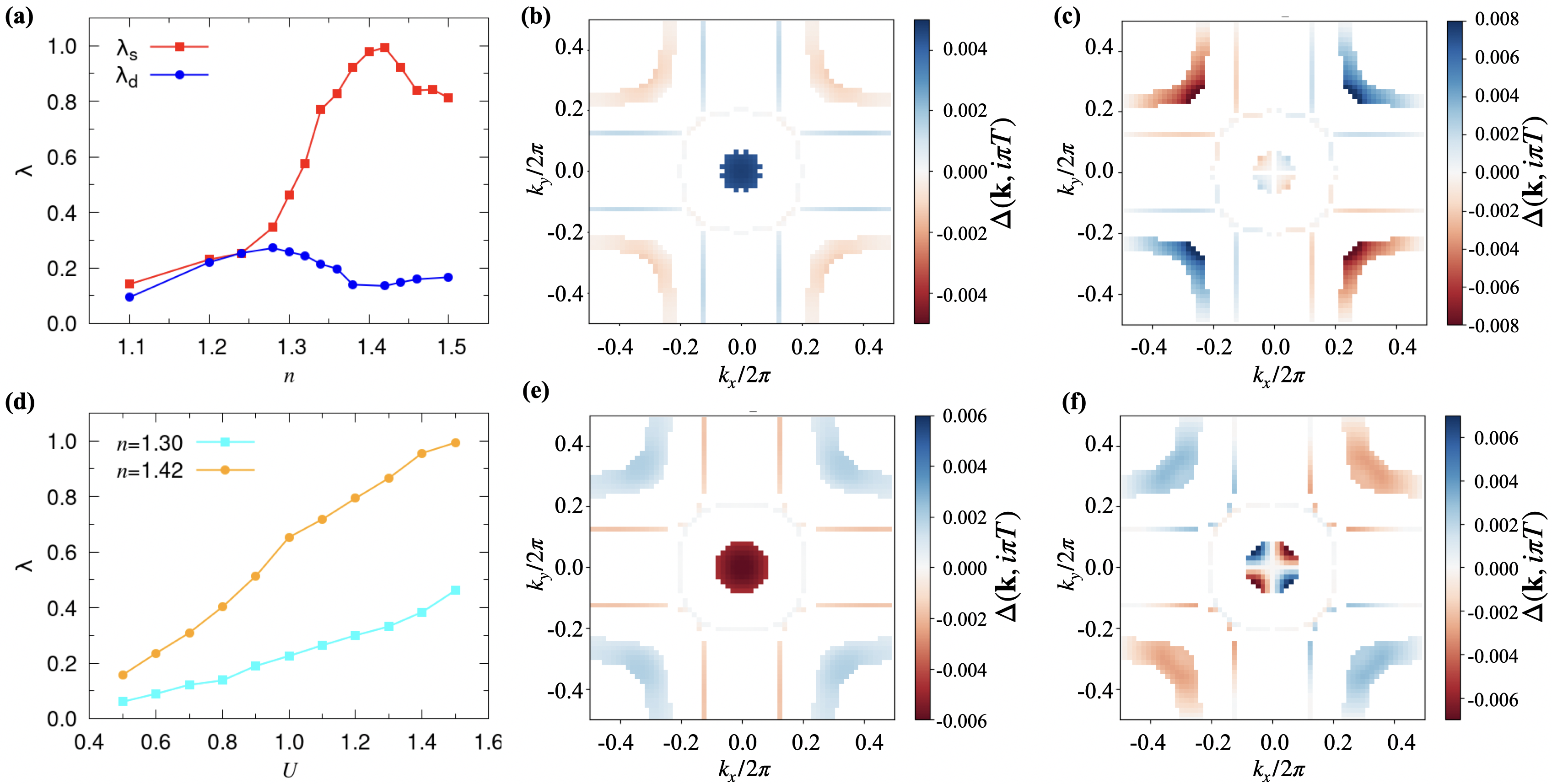}
    \caption{(a) The $\lambda$ as a function of electron filling $n$ at $U=1.5$, with the red line corresponding to $d_{xy}$-wave symmetry and the black line the $s$-wave symmetry. (b)-(c) The gap function $\Delta(\mathbf{k,i\omega_{0}})$ of the linearized Eliashberg equation within $s$- and $d_{xy}$-wave symmetry, respectively, at $n=1.3$. 
    (d) The $\lambda$ in $s$-wave symmetry as a function of $U$, with the red line corresponding to $n=1.42$ and the black line $n=1.3$.
    (e)-(f) The gap function of the linearized Eliashberg equation within $s$- and $d_{xy}$-wave symmetry , respectively, at $n=1.42$. \textcolor{black}{The gap amplitude is shown in arbitrary units after normalization in each panel.}}
    \label{fig3}
\end{figure*}

\section{Results} \ 

Since the previous computational results for the  La$_3$Ni$_2$O$_7$ superconducting film indicate that the electron filling $n$ should lie within the hole doped regime \cite{yue2025correlated}, we first choose two different hole doping levels for discussion. To ensure the reliability of FLEX, we set $U=1.5$ throughout.
\textcolor{black}{
The Fermi surface is resolved approximately by $A({\bf k},0)\approx -\mathrm{Im}\,\mathrm{Tr}\,\hat{G}(\mathbf{k},i\omega_0)/\pi$ without analytic continuation as our temperature $T=0.001$ eV is  enough low.
 The spin-susceptibility maps show the largest eigenvalue of the spin-susceptibility matrix $\hat{\chi}^{S}(\mathbf{q})$. For the gap-function plots, the color scale represents the normalized gap amplitude projected onto the band basis. Since the eigenvector of the linearized Eliashberg equation has an arbitrary overall normalization, only the relative sign and momentum dependence of the gap function are physically meaningful.} 

Fig.~\ref{fig2} (a) and (b) illustrate the self-consistent FS for $n=1.3$ and $n=1.42$. The FS is composed of three electron pockets and one hole pocket, labeled $\delta, \alpha, \beta,$ and $\gamma$ in order of increasing distance to the $\Gamma$ point. \textcolor{black}{As the electron filling $n$ increases, the $\gamma$ pocket shrinks, whereas the $\delta$ pocket around the $\Gamma$ point gradually expands. At $n=1.3$, the $\delta$ pocket is nearly absent.} Fig.~\ref{fig2}(c) and (d) show the largest eigenvalues of the spin susceptibility corresponding to $n=1.3$ and $n=1.42$, respectively. At $n=1.3$, the spin susceptibility exhibits four distinct series of peaks, which are marked in Fig.~\ref{fig2}(a). Each of these peaks corresponds to a specific nesting condition in the Fermi surface:
(1) $Q_{1}$ has two possible nesting contributions, one from the $\delta$ pocket to the $\gamma$ pocket and the other from the $\alpha$ pocket to the $\beta$ pocket.
(2) $Q_{2}$ corresponds to the nesting from the $\alpha$ pocket to the $\gamma$ pocket.
(3) $Q_{3}$ corresponds to the nesting within the $\gamma$ pocket to itself.
(4) $Q_{4}$ corresponds to the nesting from the $\beta$ pocket to the $\gamma$ pocket.
In contrast to the multi-peak structure at $n=1.3$, the spin susceptibility at $n=1.42$ is dominated by peaks at $Q_1$ and $Q_4$, with the $Q_1$ mode being particularly pronounced. This enhancement arises from two factors. First, the $\gamma$ pocket expands at $n=1.42$, and its curvature along the diagonal direction flattens, leading to a more extended tangential region with the $\beta$ pocket. Second, the nesting between the $\alpha$ and $\delta$ pockets improves due to better geometric congruence (i.e., the nested segments are nearly parallel). These factors cooperatively strengthen the spin fluctuations at $Q_1$. Concurrently, the shrinkage and increased curvature of the $\gamma$ pocket suppress the scattering channels at $Q_2$ and $Q_3$.

\begin{figure*}[t]     \includegraphics[clip,width=0.8\paperwidth,angle=0]{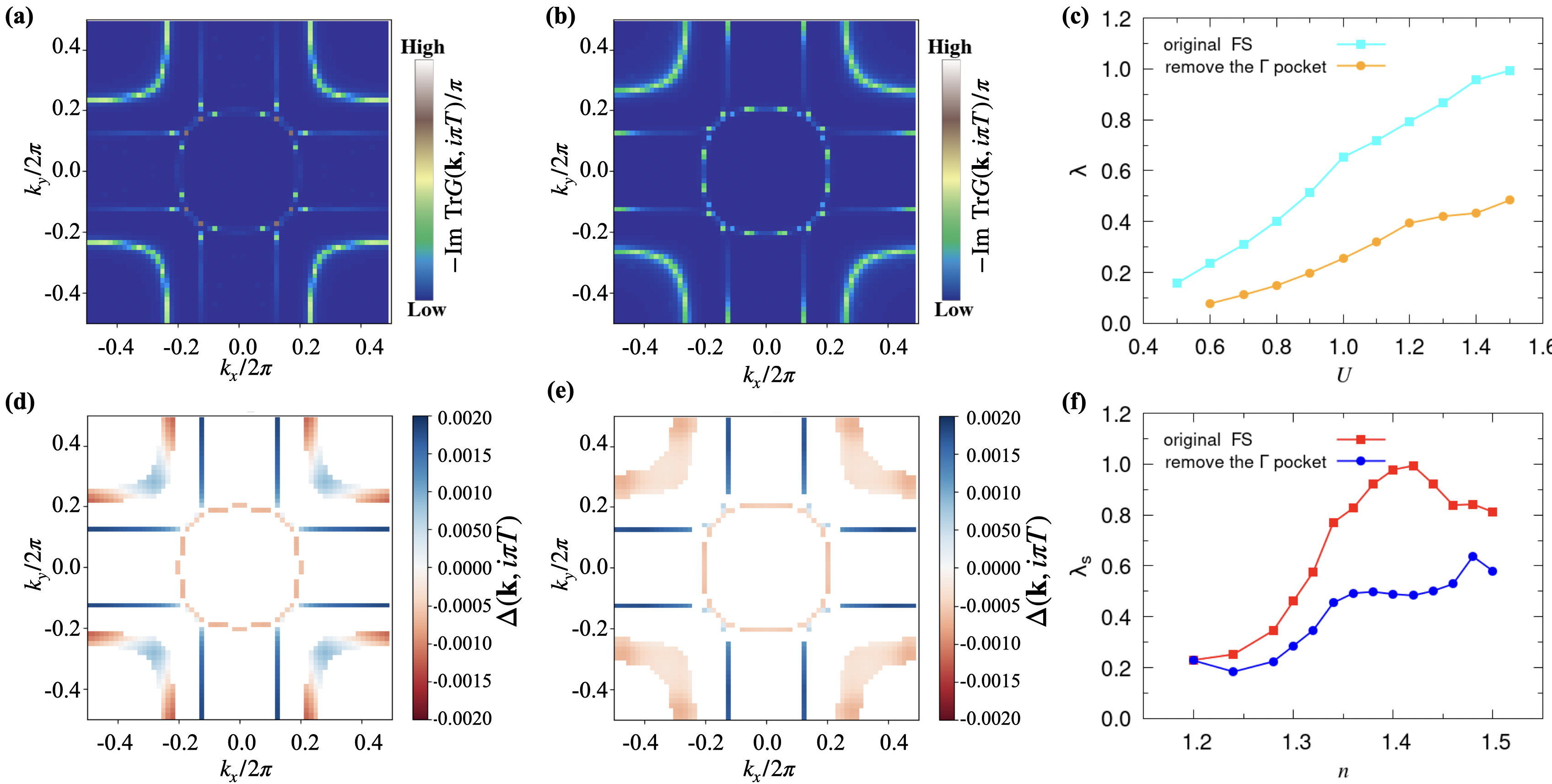}
    \caption{\textcolor{black}{
(a),(b) Low-energy spectral-weight maps used in the Eliashberg analysis
after suppressing the $\delta$-pocket contribution around the
$\Gamma$ point for $n=1.3$ and $n=1.42$, respectively. The suppression
is applied only after the FLEX self-consistent calculation has converged,
with the FLEX self-energy, susceptibilities, and pairing interaction kept
fixed. 
(d),(e) Corresponding normalized $s$-wave gap functions obtained from
the linearized Eliashberg equation. The gap amplitude is shown in arbitrary
units, and the color scale is centered at zero to indicate the relative
sign structure.
(c),(f) The $s$-wave eigenvalue $\lambda_s$ with and without the
$\delta$-pocket contribution: (c) as a function of $U$ at $n=1.42$ and
(f) as a function of electron filling $n$ at $U=1.5$.
}}
    \label{fig4}
\end{figure*}

Fig.~\ref{fig3}(a) shows the variation of the eigenvalues $\lambda$ for s-wave and d-wave solutions of the linearized Eliashberg equation as a function of electron filling $n$. \textcolor{black}{We emphasize that the eigenvalue $\lambda$ is used here as a dimensionless indicator of the relative superconducting pairing tendency at a fixed temperature $T=0.001$ eV.} It can be seen that in the over-doped hole-doped region ($n<1.3$), both s-wave and d-wave $\lambda$ values remain relatively small and exhibit competition, with the instability d-wave pairing having a slight advantage. However, in the region of lower hole doping, the s-wave eigenvalue $\lambda_{s}$ increases significantly with increasing $n$ and then forming a dome-like structure near $n = 1.42$. This phenomenon can be understood by considering the FS and spin susceptibility as shown in Fig.~\ref{fig2}, along with the gap function. Fig.~\ref{fig3}(d) shows the variation of $\lambda_s$ as a function of interaction strength $U$ for two different doping levels. As $U$ increases, $\lambda_s$ exhibits a steady growth. Fig.~\ref{fig3}(b) and (c) show the gap functions on the Fermi surface for s-wave and d-wave symmetries at $n = 1.3$, obtained after the power method converges and is already transformed into the band representation. Among the four types of nesting at this electron filling, only $Q_{3}$ (see Fig.~\ref{fig2}) occurs within the intra-band, providing spin fluctuations that favor $d_{xy}$-wave pairing. In contrast, the other three inter-band nesting processes contribute to the possibility of $s_{\pm}$-wave pairing. From the gap function analysis, although the $\delta$ pocket has almost vanished, the relatively flat band near the $\Gamma$-point still generates a noticeable signal. This suggests that in the over-doped hole-doped region, there exists a competition between $s_{\pm}$-wave pairing, primarily driven by $Q_{1}$ and $Q_{4}$, and $d_{xy}$-wave pairing, dominated by $Q_{3}$. Fig.~\ref{fig3}(e) and (f) show the gap functions for the s-wave and d-wave symmetries at $n = 1.42$, where this doping level is located at the dome of the $\lambda_{s}(n)$ which is closest to 1. Compared to the case of $n = 1.3$, the Fermi surface at this electron filling exhibits a larger density of states on the $\delta$ pocket, which matches better in size with the $\gamma$ pocket. This suggests that the significant enhancement of superconductivity at this doping level is driven by the strengthened spin fluctuations at the $Q_{1}$ wave vector, which favor pairing. However, for lower hole doping, although the shape matching between the $\delta$ pocket and the $\gamma$ pocket may improve the nesting, the density of states in the $\delta$ pocket decreases due to departure from the flat band near the $\Gamma$ point. Consequently, the spin fluctuations at $Q_{1}$ do not continue to increase, which may explain the presence of the dome structure at $\lambda_{s}(n)$.

In the previous discussion, we observed that in the weak coupling regime, the s-wave superconductivity in this system may be contributed by four types of nesting and the $\delta$ pocket appears to have a significant impact on superconductivity through the nesting with the $Q_{1}$ wave vector. To further investigate the impact of the $\delta$ pocket on superconductivity, we performed a computational experiment by artificially suppressing the Green's function spectral weight around the $\Gamma$ point in the Eliashberg calculation \textcolor{black}{after the FLEX self-consistent calculation was fully converged. Specifically, we set the Green's function near $\Gamma$ point being zero with only the $\delta$ pocket removed. Afterwards, we solve the gap function Eq.12 again.} This allows us to isolate the specific contribution of the $\delta$ pocket to the superconducting instability. After artificially removing the $\delta$ pocket, the Fermi surfaces for $n = 1.3$ and $n=1.42$ are shown in Fig.~\ref{fig4}(a) and (b). We compared $\lambda_{s}$ at $n = 1.42$ in cases where the $\delta$ pocket is present and absent, with the results shown in Fig.~\ref{fig4}(c). The gap functions after removing the $\delta$ pocket are displayed in Fig.~\ref{fig4}(d) and (e). The original $\lambda_{s}$ is consistently higher than that of the case where the $\delta$ pocket is removed for all values of $U$, indicating that although the nesting between the $\alpha$ pocket and $\beta$ pocket, as well as that between the $\gamma$ pocket and $\beta$ pocket (i.e. spin fluctuation of wave vector $Q_{1}$ and $Q_{3}$), still induce the $s_{\pm}$-wave pairing. However, in the presence of the $\delta$ pocket, the more stable $s_{\pm}$-wave pairing occurs between the $\delta$ pocket and the $\gamma$ pocket. Fig.~\ref{fig4}(f) shows the evolution of $\lambda_{s}$ with $n$ in both cases, demonstrating that this enhancement remains effective across a broad hole-doped regime. Moreover, after removing the $\delta$ pocket, the dome structure previously observed at $n = 1.42$ disappears, confirming that the optimal $T_c$ is dictated by the evolution of the $d_{z^2}$-derived Fermi surface near the $\Gamma$ point.

\section{ Discussion and Conclusions.}

In this work, we used the FLEX method to study a previously proposed model Hamiltonian for the ambient pressure La$_3$Ni$_2$O$_7$ film superconductor, which is a two-site two-orbital Hubbard model. We investigated the superconducting properties of this model in the weakly correlated, hole-doped region. We conclude that the complex spin fluctuations induced by repulsive Coulomb interactions can facilitate $s_{\pm}$-wave pairing on the $d_{z^{2}}$ orbitals, which aligns with the recent theoretical findings by Gao \cite{gao2025theoretical} \textcolor{black}{and Cao et al. \cite{Cao_2604} where the $\delta$-pocket all appears in their models}. By examining the features of the Fermi surface and correlation functions, we discussed the impact of four spin susceptibility peaks caused by five different nesting mechanisms on superconductivity. We found that a non-trivial $\delta$ pocket in the studied band structure significantly enhances the tendency toward $s_{\pm}$-wave superconductivity. Furthermore, we analyzed the consistency between the Fermi surface, spin susceptibility, superconducting instability indicator $\lambda$, and the gap function as they evolve with electron filling $n$ and interaction $U$. Finally, we confirmed this superconductivity enhancing mechanism by artificially tuning the presence or absence of the $\delta$ pocket.

\textcolor{black}{We finally comment on the possible experimental realization of the $\delta$ pocket. To the best of our knowledge, the $\delta$ pocket has not yet been directly observed in either bulk or thin-film La$_3$Ni$_2$O$_7$. 
Band engineering like electron doping \cite{LHY2507}, tensile strain or substrate substitution could possibly shift the $\delta$ band downward to the Fermi surface in new material realizations. 
Our results presented here should be viewed as a theoretical proposal for a favorable Fermi-surface topology assisted pairing: if the antibonding $d_{z^2}$ band can be tuned to be incipient or to cross the Fermi level, the associated $\delta$ pocket can cooperate with the $\gamma$ pocket and enhance the spin-fluctuation channel that favors sign-changing $s$-wave pairing.
In this sense, the $\delta$-pocket mechanism proposed here provides a possible guiding line for searching a new nickelate thin films with a higher $T_c$.}

{\it Acknowledgements. ---}
Changming Yue acknowledges support from the National Natural
Science Foundation of China (12474231), Guangdong Provincial Quantum Science Strategic Initiative (GDZX2401004). Jian-Jian Miao is supported by NSFC (No. 12404171), the Guangdong Project (Grant No. 2024QN11X176). Discussions with N. Witt and K. Kuroki are acknowledged.


\begin{thebibliography}{99}
\bibitem{LDF2019} 
D. Li, K. Lee, B.-Y. Wang, M. Osada, S. Crossley, H.-R. Lee, Y. Cui, Y. Hikita, H.-Y Hwang, Superconductivity in an infinite-layer nickelate. Nature,  572(7771): 624-627 (2019). 
\bibitem{Wang327_Nature2023} H. Sun, M. Huo, X. Hu, J. Li, Z. Liu, Y. Han, L. Tang, Z. Mao, P. Yang, B. Wang, J. Cheng, D.-X. Yao, G.-M. Zhang, and M. Wang, Signatures of superconductivity near 80 K in a nickelate under high pressure, Nature (London) 621, 493 (2023).
\bibitem{hou2023emergence} J. Hou, P.-T. Yang, Z.-Y. Liu, J.-Y. Li, P.-F. Shan, L. Ma, G. Wang, N.-N. Wang, H.-Z. Guo, J.-P. Sun, \textit{et al.}, Emergence of high-temperature superconducting phase in pressurized La$_3$Ni$_2$O$_7$ crystals, Chin. Phys. Lett. \textbf{40}, 117302 (2023).
\bibitem{Zhang_2024_HighTc} Y. Zhang, D. Su, Y. Huang, Z. Shan, H. Sun, M. Huo, K. Ye, J. Zhang, Z. Yang, Y. Xu, Y. Su, R. Li, M. Smidman, M. Wang, L. Jiao, and H. Yuan,High-temperature superconductivity with zero resistance and strange-metal behaviour in La$_3$Ni$_2$O$_{7-\delta}$, Nat. Phys. \textbf{20}, 1269--1273 (2024).
\bibitem{Xie_2024_MagneticExchange}T. Xie, M. Huo, X. Ni, F. Shen, X. Huang, H. Sun, H. C. Walker, D. Adroja, D. Yu, B. Shen, L. He, K. Cao, and M. Wang,Strong interlayer magnetic exchange coupling in La$_3$Ni$_2$O$_{7-\delta}$ revealed by inelastic neutron scattering,Sci. Bull. \textbf{69}, 3221--3227 (2024).
\bibitem{Li_2024_ElectronicCorrelation}Y. Li, X. Du, Y. Cao, C. Pei, M. Zhang, W. Zhao, K. Zhai, R. Xu, Z. Liu, Z. Li, J. Zhao, G. Li, Y. Qi, H. Guo, Y. Chen, and L. Yang,Electronic correlation and pseudogap-like behavior of high-temperature superconductor La$_3$Ni$_2$O$_7$, Chin. Phys. Lett. \textbf{41}, 087402 (2024).
\bibitem{Wang_2024_La2PrNi2O7_Nature} N. Wang, G. Wang,X. Shen, J. Hou,J. Luo, X. Ma, H. Yang, L. Shi, J. Dou, J. Feng, J. Yang, Y. Shi, Z. Ren, H. Ma, P. Yang, Z. Liu, Y. Liu, H. Zhang, X. Dong, Y. Wang, K. Jiang, J. Hu, S. Nagasaki, K. Kitagawa, S. Calder, J. Yan, J. Sun, B. Wang, R. Zhou, Y. Uwatoko, J. Cheng, Bulk high-temperature superconductivity in pressurized tetragonal La$_2$PrNi$_2$O$_7$, Nature \textbf{634}, 579-584 (2024).
\bibitem{dan_2024}
D. Zhao, Y. Zhou, M. Huo, Y. Wang, L. Nie, Y. Yang, J. Ying, M. Wang, T. Wu, and X. Chen,
Pressure-enhanced spin-density-wave transition in double-layer nickelate La$_3$Ni$_2$O$_{7-\delta}$,
\textit{Science Bulletin} (2025).

\bibitem{wang2024pressure} G. Wang, N. N. Wang, X. L. Shen, J. Hou, L. Ma, L. F. Shi, Z. A. Ren, Y. D. Gu, H. M. Ma, P. T. Yang, \textit{et al.}, Pressure-Induced Superconductivity in Polycrystalline La$_3$Ni$_2$O$_{7-\delta}$, Phys. Rev. X \textbf{14}, 011040 (2024).
\bibitem{cai_2024}
S. Cai, Y. Zhou, H. Sun, K. Zhang, J. Zhao, M. Huo, L. Nataf, Y. Wang, J. Li, J. Guo, \textit{et al.},
Low-temperature mean valence of nickel ions in pressurized La$_3$Ni$_2$O$_7$,
Phys. Rev. B \textbf{111}, 104511 (2025).
\bibitem{ren2025resolving}
X. Ren, R. Sutarto, X. Wu, J. Zhang, H. Huang, T. Xiang,
J. Hu, R. Comin, X. Zhou, and Z. Zhu,
Resolving the electronic ground state of La$_3$Ni$_2$O$_7$-$\delta$ films,
\textit{Commun. Phys.} \textbf{8}, 52 (2025).

\bibitem{liu_2024} Z. Liu, M. Huo, J. Li, Q. Li, Y. Liu, Y. Dai, X. Zhou, J. Hao, Y. Lu, M. Wang, \textit{et al.}, Electronic correlations and partial gap in the bilayer nickelate La$_3$Ni$_2$O$_7$, Nat. Commun. \textbf{15}, 7570 (2024).






\bibitem{YaoPRL2023} Zhihui Luo, Xunwu Hu, Meng Wang, Wei Wu, and Dao-Xin Yao, Bilayer Two-Orbital Model of La$_3$Ni$_2$O$_7$ under Pressure, Physical Review Letters \textbf{131}, 126001 (2023).
\bibitem{Yang_2023_sPlusMinus} Q.-G. Yang, D. Wang, and Q.-H. Wang, Possible $s^\pm$-wave superconductivity in La$_3$Ni$_2$O$_7$, Phys. Rev. B \textbf{108}, L140505 (2023).
\bibitem{Liu_2023_sPlusMinus} Y.-B. Liu, J.-W. Mei, F. Ye, W.-Q. Chen, and F. Yang, $s^\pm$-Wave Pairing and the Destructive Role of Apical-Oxygen Deficiencies in La$_3$Ni$_2$O$_7$ under Pressure, Phys. Rev. Lett. \textbf{131}, 236002 (2023).
\bibitem{Yang_2023_ValenceBonds} Y.-F. Yang, G.-M. Zhang, and F.-C. Zhang, Interlayer valence bonds and two-component theory for high-$T_c$ superconductivity of La$_3$Ni$_2$O$_7$ under pressure, Phys. Rev. B \textbf{108}, L201108 (2023).
\bibitem{Qin_2023_HighTc} Q. Qin and Y.-F. Yang, High-$T_c$ superconductivity by mobilizing local spin singlets and possible route to higher $T_c$ in pressurized La$_3$Ni$_2$O$_7$, Phys. Rev. B \textbf{108}, L140504 (2023).
\bibitem{zhang_2023} Y. Zhang, L.-F. Lin, A. Moreo, and E. Dagotto, Electronic structure, dimer physics, orbital-selective behavior, and magnetic tendencies in the bilayer nickelate superconductor La$_3$Ni$_2$O$_7$ under pressure, Phys. Rev. B \textbf{108}, L180510 (2023).
\bibitem{Christiansson_2023}Viktor Christiansson, Francesco Petocchi, and Philipp Werner, Correlated Electronic Structure of La$_3$Ni$_2$O$_7$ under Pressure, Phys. Rev. Lett. \textbf{131}, 206501 (2023).
\bibitem{pupha_l2023}
P. Puphal, P. Reiss, N. Enderlein, Y.-M. Wu, G. Khaliullin, V. Sundaramurthy, T. Priessnitz, M. Knauft, A. Suthar, L. Richter, \textit{et al.},
Unconventional crystal structure of the high-pressure superconductor La$_3$Ni$_2$O$_7$,
Phys. Rev. Lett. \textbf{133}, 146002 (2024).
\bibitem{liao2023electron} Z. Liao, L. Chen, G. Duan, Y. Wang, C. Liu, R. Yu, and Q. Si,  
Electron correlations and superconductivity in La$_3$Ni$_2$O$_7$ under pressure tuning,  
Phys. Rev. B \textbf{108}, 214522 (2023).
\bibitem{liao_2024} Z. Liao, Y. Wang, L. Chen, G. Duan, R. Yu, and Q. Si, Orbital-selective electron correlations in high-$T_c$ bilayer nickelates: from a global phase diagram to implications for spectroscopy, arXiv:2412.21019 (2024).
\bibitem{zhang2024electronic} Y. Zhang, L.-F. Lin, A. Moreo, T. A. Maier, and E. Dagotto, Electronic structure, self-doping, and superconducting instability in the alternating single-layer trilayer stacking nickelates La$_3$Ni$_2$O$_7$, Phys. Rev. B \textbf{110}, L060510 (2024).
\bibitem{zhang2023trends} Y. Zhang, L.-F. Lin, A. Moreo, T. A. Maier, and E. Dagotto, Trends in electronic structures and $s\pm$-wave pairing for the rare-earth series in bilayer nickelate superconductor R$_3$Ni$_2$O$_7$, Phys. Rev. B \textbf{108}, 165141 (2023).
\bibitem{Tian_2024_Correlation} Y.-H. Tian, Y. Chen, J.-M. Wang, R.-Q. He, and Z.-Y. Lu, Correlation effects and concomitant two-orbital $s^\pm$-wave superconductivity in La$_3$Ni$_2$O$_7$ under high pressure, Phys. Rev. B \textbf{109}, 165154 (2024).
\bibitem{Oh_2024_StrongPairing} H. Yang, H. Oh, and Y.-H. Zhang, Strong pairing from a small Fermi surface beyond weak coupling: Application to La$_3$Ni$_2$O$_7$, Phys. Rev. B \textbf{110}, 104517 (2024).
\bibitem{Fan_2024_BilayerCoupling} Z. Fan, J.-F. Zhang, B. Zhan, D. Lv, X.-Y. Jiang, B. Normand, and T. Xiang, \textit{Superconductivity in nickelate and cuprate superconductors with strong bilayer coupling}, Phys. Rev. B \textbf{110}, 024514 (2024).
\bibitem{RyeePRL2024CMDFT} Siheon Ryee, Niklas Witt, and Tim O. Wehling, Quenched Pair Breaking by Interlayer Correlations as a Key to Superconductivity in La$_3$Ni$_2$O$_7$, Phys. Rev. Lett. \textbf{133}, 096002 (2024).
\bibitem{wang2024electronic}
Y. Wang, K. Jiang, Z. Wang, F.-C. Zhang, and J. Hu,
Electronic and magnetic structures of bilayer La$_3$Ni$_2$O$_7$ at ambient pressure,
Phys. Rev. B \textbf{110}, 205122 (2024).
\bibitem{chen2024electronic} X. Chen, J. Choi, Z. Jiang, J. Mei, K. Jiang, J. Li, S. Agrestini, M. Garcia-Fernandez, H. Sun, X. Huang, \textit{et al.}, Electronic and magnetic excitations in La$_3$Ni$_2$O$_7$, Nat. Commun. \textbf{15}, 9597 (2024).
\bibitem{lu2024interlayer} C. Lu, Z. Pan, F. Yang, and C. Wu, Interlayer-coupling-driven high-temperature superconductivity in La$_3$Ni$_2$O$_7$ under pressure, Phys. Rev. Lett. \textbf{132}, 146002 (2024).
\bibitem{sakakibara2024} H. Sakakibara, N. Kitamine, M. Ochi, and K. Kuroki, Possible high $T_c$ superconductivity in La$_3$Ni$_2$O$_7$ under high pressure through manifestation of a nearly half-filled bilayer Hubbard model, Phys. Rev. Lett. \textbf{132}, 106002 (2024).
\bibitem{jiang2024high} K. Jiang, Z. Wang, and F.-C. Zhang, High-temperature superconductivity in La$_3$Ni$_2$O$_7$, Chin. Phys. Lett. \textbf{41}, 017402 (2024).
\bibitem{zhang2024strong} J.-X. Zhang, H.-K. Zhang, Y.-Z. You, and Z.-Y. Weng, Strong pairing originated from an emergent $Z_2$ Berry phase in La$_3$Ni$_2$O$_7$, Phys. Rev. Lett. \textbf{133}, 126501 (2024).
\bibitem{Zheng_2025_TwoOrbital} Y.-Y. Zheng and W. Wu, $s^\pm$-wave superconductivity in the bilayer two-orbital Hubbard model, Phys. Rev. B \textbf{111}, 035108 (2025).
\bibitem{bejas2025out}
M. Bejas, X. Wu, D. Chakraborty, A. P. Schnyder, and A. Greco,
Out-of-plane bond-order phase, superconductivity, and their competition in the
$t$-$J_{\parallel}$-$J_{\perp}$ model: Possible implications for bilayer nickelates,
\textit{Phys. Rev. B} \textbf{111}, 144514 (2025).
\bibitem{puphal2025superconductivity}
P. Puphal, T. Sch{\"a}fer, B. Keimer, and M. Hepting,
Superconductivity in infinite-layer and Ruddlesden--Popper nickelates,
\textit{Nat. Rev. Phys.} 1--16 (2025).
\bibitem{jiang2025theory} K.-Y. Jiang, Y.-H. Cao, Q.-G. Yang, H.-Y. Lu, and Q.-H. Wang,  
Theory of pressure dependence of superconductivity in bilayer nickelate La$_3$Ni$_2$O$_7$,  
Physical Review Letters. \textbf{134}  076001 (2025).
\bibitem{duan2025orbital} G. Duan, Z. Liao, L. Chen, Y. Wang, R. Yu, and Q. Si, 
Orbital-selective correlation effects and superconducting pairing symmetry in a multiorbital $t$-$J$ model for bilayer nickelates, arXiv:2502.09195 (2025).
\bibitem{jiang2025dual}
P. Jiang, J. Li, Y.-H. Cao, X. Cao, Z. Zhong, Y. Lu, and Q.-H. Wang,
Dual instability of superconductivity from oxygen defects in La$_3$Ni$_2$O$_{7+\delta}$,
arXiv:2512.00301 (2025).
\bibitem{xi2025transition} W. Xi, S.-L. Yu, and J.-X. Li, Transition from $s\pm$-wave to $d_{x^2-y^2}$-wave superconductivity driven by interlayer interaction in the bilayer two-orbital model of La$_3$Ni$_2$O$_7$, 
Phys. Rev. B \textbf{111}, 104505 (2025).
\bibitem{shi2025theoretical} L. Shi, Y. Luo, W. Wu, and Y. Zhang, Theoretical Investigation of High-Tc Superconductivity in Sr-Doped La$_3$Ni$_2$O$_7$ at Ambient Pressure,  arXiv preprint arXiv:2503.13197 (2025).
\bibitem{ouyang2025two} Z. Ouyang, R.-Q. He, and Z.-Y. Lu,  Two key factors to superconductivity of Ruddlesden-Popper nickelates: enhanced quasi-particle weight and strong local spin fluctuation,  arXiv preprint arXiv:2503.08682 (2025).
\bibitem{xia2025sensitive}
C. Xia, H. Liu, S. Zhou, and H. Chen,
Sensitive dependence of pairing symmetry on Ni-$e_g$ crystal field splitting in the nickelate superconductor La$_3$Ni$_2$O$_7$,
\textit{Nat. Commun.} \textbf{16}, 1054 (2025).
\bibitem{maier2025interlayer}
T. A. Maier, P. Doak, L.-F. Lin, Y. Zhang, A. Moreo, and E. Dagotto,
Interlayer pairing in bilayer nickelates,
\textit{npj Quantum Materials} (2026).

\bibitem{wang2025electronic} Wang Y, Zhang Y, Jiang K. Electronic structure and disorder effect of La$_3$Ni$_2$O$_7$ superconductor[J]. Chinese Physics B, 2025, 34(4): 047105.

\bibitem{wu2024ac}
S. Wu, Z. Yang, X. Ma, J. Dai, M. Shi, H.-Q. Yuan, H.-Q. Lin, and C. Cao,
Ac$_3$Ni$_2$O$_7$ and La$_2$AeNi$_2$O$_6$F ($Ae$ = Sr, Ba): Benchmark materials for bilayer nickelate superconductivity,
arXiv:2403.11713 (2024).
\bibitem{huo2025modulation} Huo Z, Luo Z, Zhang P, Yang A, Liu Z, Tao X, Zhang Z, Guo S, Jiang Q, Chen W, et al. Modulation of the octahedral structure and potential superconductivity of La$_3$Ni$_2$O$_7$ through strain engineering[J]. Science China Physics, Mechanics Astronomy, 2025, 68(3): 237411 (2025).
\bibitem{wang2023observation} Wang G, Wang N, Wang Y, Shi L, Shen X, Hou J, Ma H, Yang P, Liu Z, Zhang H, et al. Observation of high-temperature superconductivity in the high-pressure tetragonal phase of La$_{2}$PrNi$_{2}$O$_{7}$-$_{\delta}$[J]. arXiv preprint arXiv:2311.08212, 2023.
\bibitem{rhodes2024structural} Rhodes L C, Wahl P. Structural routes to stabilize superconducting La$_3$Ni$_2$O$_7$ at ambient pressure[J]. Physical Review Materials, 2024, 8(4): 044801.
\bibitem{geisler2024structural} Geisler B, Hamlin J J, Stewart G R, Hennig R G, Hirschfeld P J. Structural transitions, octahedral rotations, and electronic properties of A$_{3}$Ni$_{2}$O$_{7}$ rare-earth nickelates under high pressure[J]. npj Quantum Materials, 2024, 9(1): 38.
\bibitem{chen2025charge} Chen X, Jiang P, Li J, Zhong Z, Lu Y. Charge and spin instabilities in superconducting La$_3$Ni$_2$O$_7$[J]. Physical Review B, 2025, 111(1): 014515 (2025).



\bibitem{Ko_LNO327AmbientSC} E. Ko, Y. Yu, Y. Liu, L. Bhatt, J. Li, V. Thampy, C. Kuo, B. Wang, Y. Lee, K. Lee, J. Lee, B. H. Goodge, D. A. Muller, H. Y. Hwang, Signatures of ambient pressure superconductivity in thin film La$_3$Ni$_2$O$_7$, Nature (2024).
\bibitem{Zhou_2024_AmbientSC}
G. Zhou, W. Lv, H. Wang, Z. Nie, Y. Chen, Y. Li, H. Huang, W.-Q. Chen, Y.-J. Sun, Q.-K. Xue, \textit{et al.},
Ambient-pressure superconductivity onset above 40 K in (La, Pr)$_3$Ni$_2$O$_7$ films,
\textit{Nature} \textbf{640}, 641--646 (2025).
\bibitem{zhou2025superconductivity}
G. Zhou, H. Wang, H. Huang, Y. Chen, F. Peng, W. Lv, Z. Nie, W. Wang, Q.-K. Xue, and Z. Chen, Superconductivity onset above 60 K in ambient-pressure nickelate films, arXiv:2512.04708 (2025).
\bibitem{wang2026superconducting}
M. Wang, B. Hao, W. Sun, S. Yan, S. Sun, H. Zhang, Z. Gu, and Y. Nie,
Superconducting dome in La$_{3-x}$Sr$_x$Ni$_2$O$_{7-\delta}$ thin films,
Phys. Rev. Lett. \textbf{136}, 066002 (2026).
\bibitem{ryee2025superconductivity}
S. Ryee, N. Witt, G. Sangiovanni, and T. O. Wehling,
Superconductivity governed by Janus-faced Fermiology in strained bilayer nickelates,
arXiv:2506.21480 (2025).
\bibitem{ushio2025theoretical}
K. Ushio, S. Kamiyama, Y. Hoshi, R. Mizuno, M. Ochi, K. Kuroki, and H. Sakakibara,
Theoretical study on ambient pressure superconductivity in La$_3$Ni$_2$O$_7$ thin films: structural analysis, model construction, and robustness of $s\pm$-wave pairing,
arXiv:2506.20497 (2025).
\bibitem{yue2025correlated}
C. Yue, J.-J. Miao, H. Huang, Y. Hua, P. Li, Y. Li, G. Zhou, W. Lv, Q. Yang, F. Yang, \textit{et al.},
Correlated electronic structures and unconventional superconductivity in bilayer nickelate heterostructures,
\textit{Natl. Sci. Rev.}, nwaf253 (2025).
\bibitem{gao2025theoretical} Y. Gao, Theoretical study of the electronic correlation and superconducting pairing in La$_{2.85}$Pr$_{0.15}$Ni$_2$O$_7$ film grown on SrLaAlO$_4$), arXiv:2507.19784 (2025).
\bibitem{shi2025effect}
H. Shi, Z. Huo, G. Li, H. Ma, T. Cui, D. Yao, and D. Duan,
The effect of carrier doping and thickness on the electronic structures of La$_3$Ni$_2$O$_7$ thin films,
Chin. Phys. Lett. \textbf{42}, 080708 (2025).
\bibitem{le2025opposite}
C. Le, J. Zhan, X. Wu, and J. Hu,
Opposite-mirror-parity scattering as the origin of superconductivity in strained
bilayer nickelates,
arXiv:2501.14665 (2025).
\bibitem{liu2025superconductivity}
Y. Liu, E. K. Ko, Y. Tarn, L. Bhatt, J. Li, V. Thampy, B. H. Goodge, D. A. Muller, S. Raghu, Y. Yu, \textit{et al.},
Superconductivity and normal-state transport in compressively strained La$_2$PrNi$_2$O$_7$ thin films,
\textit{Nature Materials}, 1--7 (2025).
\bibitem{XJZhou_ARPES_327} J. Yang, H. Sun, X. Hu, Y. Xie, T. Miao, H. Luo, H. Chen, B. Liang, W. Zhu, G. Qu, C.-Q. Chen, M. Huo, Y. Huang, S. Zhang, F. Zhang, F. Yang, Z. Wang, Q. Peng, H. Mao, G. Liu, Z. Xu, T. Qian, D.-X. Yao, M. Wang, L. Zhao, X. J. Zhou, Orbital-dependent electron correlation in double-layer nickelate La$_3$Ni$_2$O$_7$, Nat. Commun. \textbf{15}, 4373 (2024).
\bibitem{li2025photoemission}
P. Li, G. Zhou, W. Lv, Y. Li, C. Yue, H. Huang, L. Xu, J. Shen, Y. Miao, W. Song, \textit{et al.},
Angle-resolved photoemission spectroscopy of superconducting (La, Pr)$_3$Ni$_2$O$_7$/SrLaAlO$_4$ heterostructures,
\textit{Natl. Sci. Rev.}, nwaf205 (2025).
\bibitem{osada2025strain}
M. Osada, C. Terakura, A. Kikkawa, M. Nakajima, H.-Y. Chen, Y. Nomura, Y. Tokura, and A. Tsukazaki,
Strain-tuning for superconductivity in La$_3$Ni$_2$O$_7$ thin films,
\textit{Commun. Phys.} \textbf{8}, 251 (2025).

\bibitem{hao2025superconductivity}
B. Hao, M. Wang, W. Sun, Y. Yang, Z. Mao, S. Yan, H. Sun, H. Zhang, L. Han, Z. Gu, \textit{et al.},
Superconductivity in Sr-doped La$_3$Ni$_2$O$_7$ thin films,
\textit{Nature Materials} \textbf{24}, 1756--1762 (2025).

\bibitem{wang2025recent}
Y. Wang, K. Jiang, J. Ying, T. Wu, J. Cheng, J. Hu, and X. Chen,
Recent progress in nickelate superconductors,
\textit{Natl. Sci. Rev.} \textbf{12}, nwaf373 (2025).

\bibitem{zhan2025cooperation}
J. Zhan, Y. Gu, X. Wu, and J. Hu,
Cooperation between electron-phonon coupling and electronic interaction in
bilayer nickelates La$_3$Ni$_2$O$_7$,
\textit{Phys. Rev. Lett.} \textbf{134}, 136002 (2025).

\bibitem{gu2025effective}
Y. Gu, C. Le, Z. Yang, X. Wu, and J. Hu,
Effective model and pairing tendency in the bilayer Ni-based superconductor
La$_3$Ni$_2$O$_7$,
\textit{Phys. Rev. B} \textbf{111}, 174506 (2025).
\bibitem{zhang2025mirror}
Z. Zhang, J. Zhan, C. Le, H. C. Po, J. Hu, and X. Wu,
Mirror-selective quasiparticle interference in bilayer nickelate superconductor,
arXiv:2512.14544 (2025).






\bibitem{bickers1989conserving} N. E. Bickers and D. J. Scalapino, 
Conserving approximations for strongly fluctuating electron systems. I. Formalism and calculational approach, Ann. Phys. \textbf{193}, 206--251 (1989).
\bibitem{bickers1991conserving} N. E. Bickers and S. R. White, Conserving approximations for strongly fluctuating electron systems. II. Numerical results and parquet extension, Phys. Rev. B \textbf{43}, 8044 (1991).
\bibitem{yanase2003theory} Y. Yanase, T. Jujo, T. Nomura, H. Ikeda, T. Hotta, and K. Yamada, 
Theory of superconductivity in strongly correlated electron systems, Phys. Rep. \textbf{387}, 1--149 (2003).
\bibitem{kuroki2008unconventional} K. Kuroki, S. Onari, R. Arita, H. Usui, Y. Tanaka, H. Kontani, and H. Aoki, Unconventional Pairing Originating from the Disconnected Fermi Surfaces of Superconducting LaFeAsO$_{1-x}$F$_x$, Phys. Rev. Lett. \textbf{101}, 087004 (2008).
\bibitem{kang2011spin}
J. Kang, S.-L. Yu, Z.-J. Yao, and J.-X. Li,
Spin-fluctuation-mediated pairing symmetry on the metallic kagome lattice,
\textit{J. Phys.: Condens. Matter} \textbf{23}, 175702 (2011).


\bibitem{witt2021efficient} N. Witt, E. G. C. P. Van Loon, T. Nomoto, R. Arita, and T. O. Wehling, 
Efficient fluctuation-exchange approach to low-temperature spin fluctuations and superconductivity: From the Hubbard model to Na$_x$CoO$_2${\textperiodcentered}$y$H$_2$O, Phys. Rev. B \textbf{103}, 205148 (2021).
\bibitem{bjornson2021orbital} K. Bj{\"o}rnson, A. Kreisel, A. T. R{\o}mer, and B. M. Andersen,  
Orbital-dependent self-energy effects and consequences for the superconducting gap structure in multiorbital correlated electron systems, Phys. Rev. B \textbf{103}, 024508 (2021).

\bibitem{shinaoka2017compressing}
H. Shinaoka, J. Otsuki, M. Ohzeki, and K. Yoshimi,
Compressing Green's function using intermediate representation between imaginary-time and real-frequency domains,
\textit{Phys. Rev. B} \textbf{96}, 035147 (2017).
\bibitem{wallerberger2023sparse}
M. Wallerberger, S. Badr, S. Hoshino, S. Huber, F. Kakizawa,
T. Koretsune, Y. Nagai, K. Nogaki, T. Nomoto, H. Mori, \textit{et al.},
sparse-ir: Optimal compression and sparse sampling of many-body propagators,
\textit{SoftwareX} \textbf{21}, 101266 (2023).


\bibitem{LHY2507} Y.-H. Cao ,K.-Y. Jiang ,H.-Y. Lu, D. Wang, Q.-H. Wang, arXiv:2507.13694 (2025). 

\bibitem{Cao_2604} Y.-H. Cao, K.-Y. Jiang, H.-Y. Lu, D. Wang, Q.-H. Wang arXiv:2604.05590 (2026). 
\end{thebibliography}
\end{document}